\documentclass{ws-procs10x7}
\usepackage{balance,graphicx}

\newcolumntype{d}[1]{D{.}{.}{#1}}

\def\Journal#1#2#3#4{{\it #1} {\bf #2}, #3 (#4)}

\makeindex
\begin{document}

\title{Magnetic Impurity States and Ferromagnetic Interaction 
in Diluted Magnetic Semiconductors}

\author{M. ICHIMURA$^{*,1,2}$, K. TANIKAWA$^1$, S. TAKAHASHI$^1$, 
G. BASKARAN$^{1,3}$, \and {S.~MAEKAWA$^{1,4}$}}

\address{$^1$Institute for Materials Research, Tohoku University, Sendai, 
980-8577, Japan\\$^*$E-mail: ichimura@imr.tohoku.ac.jp}
\address{$^2$Advanced Research Laboratory, Hitachi, Ltd., Hatoyama, Saitama, 
350-0395, Japan}
\address{$^3$Institute of Mathematical Sciences, C.I.T. Campus, Chennai 600 113, India}
\address{$^4$CREST, Japan Science and Technology Agency, Kawaguchi, Saitama, 
332-0012, Japan}


\twocolumn
[\maketitle\abstract
{
The electronic structure of diluted magnetic semiconductors is studied, 
especially focusing on the hole character. 
The Haldane-Anderson model is extend to a magnetic impurity, 
and is analyzed in the Hartree-Fock approximation. 
Due to the strong hybridization of an impurity $d$-states with host $p$-states, 
it is shown that holes are created in the valence band, 
at the same time the localized magnetic moments are formed. 
These holes are weakly bound around the impurity site 
with the characteristic length of several lattice constants. 
For the case of the two impurities, 
it is found that when the separation of two impurities 
is of the order of the length of holes, 
the overlap of these holes favors the ferromagnetic interaction. 
The spatially extended holes play an essential role 
for the ferromagnetic exchange interaction. 
}
]

\section{Introduction}
The spin dependent transport has attracted great interest  
from the technological point of view as well as 
the physical point of view~\cite{MaekawaShinjo}. 
For the requirement to control and manipulate the spin transport 
by external perturbations, including applied electric and/or magnetic 
fields and optical excitation, 
the diluted magnetic semiconductors (DMS) are considered 
to be the most important candidates for such promising properties. 
However, we have not been able to identify the DMS 
with Curie temperature ($T_c$) above room temperature, 
and there exists great discrepancy among the experiments for 
mechanism of ferromagnetic ground state in DMS. 

In various types of DMS, (III,Mn)V family, 
(Ga,Mn)As~\cite{Ohno1} has been extensively studied both from the 
experimental and theoretical sides. 
Matsukura {\it et al}. have shown that the carrier in (Ga,Mn)As 
is $p$-type holes by Hall effect~\cite{Matsukura}. 
These holes are considered to play an important role for 
the ferromagnetism in (Ga,Mn)As. 
Ohno {\it et al}. have realized to control the ferromagnetic properties 
such as $T_c$~\cite{Ohno2} 
and coercive field ($H_c$)~\cite{Ohno3} 
by the control of hole density in (Ga,Mn)As 
with the field-effect transistor (FET) device. 
These results have made it clear that 
(Ga,Mn)As is hole-mediated ferromagnet. 

There are many experimental studies for hole states in (Ga,Mn)As. 
The Mn $2p$ core level photoemission for (Ga,Mn) 
with the configuration interaction analysis 
has suggested that a Mn ion is the neutral (Mn)$^{+3}$ 
with $d^5+1hole$ configuration 
and that holes are localized at the Mn site~\cite{Okabayashi1}. 
On the other hand, the valence band photoemission experiment 
has shown that holes are occupied in the host valence band (VB) 
just below the Fermi energy ($E_F$)~\cite{Okabayashi2}. 
The experiment of angle resolved photoemission spectra (ARPES) 
have shown the existence of impurity band below $E_F$ without 
dispersion in the direction along $\vec{k}=(001)$~\cite{Okabayashi3}. 
The infra-red (IR) absorption experiments have shown 
the hopping conduction due to the localized holes, 
and spectral transfer from interband to intraband~\cite{Hirakawa}. 
This is interpreted that the localized holes do not provide 
the RKKY interaction between magnetic moments, 
and that the spectral transfer is caused by 
the double exchange (DE) interaction. 
The X-ray absorption spectral (XAS) experiment 
has shown that the holes are mainly in As $4p$ orbital~\cite{Ishikawa}. 
There exists disagreement over the interpretation of 
the hole character, localized or delocalized. 
%
\section{Model description}
We start with the Haldane-Anderson model, 
which has treated the charge state of the impurity $d$-state 
in semiconductor hosts, 
and has succeeded to describe the multiple charge state~\cite{H&A}. 
Here, we extend the Haldane-Anderson model to a magnetic impurity, 
and simplify the model with no orbital degeneracy. 
The  degenerate-orbital case is discussed in the following section. 
A standard Hartree-Fock (HF) approximation applied to the Haldane-Andersonas 
leads to 
\begin{eqnarray}
   \mathcal{H}_{HF} 
& = & 
   \sum_{\sigma}E^{ef\!f}_{d\sigma} n_{\sigma}
   +
   \sum_{k\sigma}\epsilon_k c_{k\sigma}^{\dagger}c_{k\sigma}\nonumber \\[4pt]
&&{} + 
   V\sum_{k\sigma}(c_{k\sigma}^{\dagger}d_{\sigma} + h.c.), \label{eq:HF-H}
\end{eqnarray}
where
\begin{equation}
   E^{ef\!f}_{d\sigma}
   = E_d
   +
   U \langle n_{-\sigma} \rangle. \label{eq:HF-E} 
\end{equation}
Here $n_{\sigma}=d_{\sigma}^{\dagger}d_{\sigma}$, 
and $c_{k\sigma}^{\dagger}$ and $d_{\sigma}^{\dagger}$ are 
the creation operators for the $p$-state of host semiconductor energy band 
and for $d$-state at impurity site, respectively. 
We assume that in the host band structure, 
the band widths of the VB and conduction band (CB) is semi-infinite 
and the density of states (DOS) of both bands are constant ($\rho_0$), 
since the interested energy region is 
near the CB bottom ($\varepsilon_c$) and VB top ($\varepsilon_v$). 
$U$ and $V$ represent the onsite Coulomb energy and 
the hybridization between $d$- and $p$-states, respectively. 
It is convenient to define $\Delta=\pi|V|^2\rho_0$. 
$E^{ef\!f}_{d\sigma}$ is the effective energy level of $d$-state. 
In this work, we restrict ourselves to the symmetric case, $E_d=-(U-\Delta_G)/2$. 
$\Delta_G$ is the gap in the host energy band. 

\begin{figure}[b]
\centerline{\includegraphics[width=1.2in]{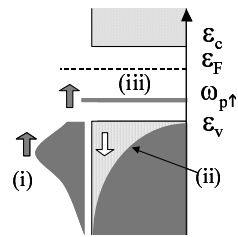}}
\caption{The schematic picture of the single particle spectrum for up-spin.
(i) DOS of VBS due to $p$-$d$ hybridization.  
(ii) DOS of host $p$-band. The DOS deviates from rectangular shape, 
corresponding to the formation of VBS. 
This deviation is regarded as the $p$-hole with down-spin. 
(iii) Near the VB top in the gap, the split-off state appears at $\omega_p$, 
which mainly consists of the host $p$-state.}
\label{fig1}
\end{figure}

The HF single particle spectrum is obtained 
as shown schematically in Fig.~\ref{fig1}.
It is shown analytically that the Friedel sum rule is satisfied 
between virtual bound states (VBS) and $p$-hole, 
and between $d$- and $p$-components in the split-off state. 
The number of states in VBS is equal to the one in $p$-hole state. 
The sum of the $d$- and $p$-components in the split-off state is always 1. 
This means that the split-off state carries the unit charge $e$ and spin $1/2$. 
These two sum rules are convenient to consider the spatial distribution of $p$-hole spin. 

In the situation of Fig.~\ref{fig1}, we assume the deep $d$-level in VB, 
{\it i.e.} negatively large $E_d$. 
Under this assumption, when the split-off state is empty, 
the electron configuration is $d^1 + 1hole$. 
When we set the split-off state to be singly occupied, 
the $d^1$ electron configuration is realized. 

By using the above relation, we can estimate the spatial distributions of 
the charge and spin for the cases of $d^1 + 1hole$ and $d^1$. 
For $d^1 + 1hole$, the charge and spin distributions are determined by $p$-hole, 
since the $d$-state is well localized like $\delta$-function. 
We obtained that the distributions of $p$-holes 
spread over several lattice constants, 
then the induced magnetic moment in host is coupled antiferromagnetically 
with the impurity spin. 
For $d^1$, the distribution of $p$-hole is cancelled by 
the occupation of the split-off state. 
The $p$-component density of the split-off state 
is calculated analytically as 
$\sim \rho_0 \Delta \exp(-2r/\xi_p)/(r/\xi_p)$, 
where $\xi_p=1/\sqrt{2m^*\omega_p}$ 
and $m^*$ is the effective mass of host. 
From this result, for $d^1 + 1hole$, it is understood that 
the $p$-hole is weakly bound around the impurity spin 
with the spreading of several lattice constants. 
%
\section{Result and Discussion}
%
Here, we extend the above result to the impurity treatment in (Ga,Mn)As. 
For this purpose, the three degenerate $d$-orbitals are considered as impurity states 
since the tetrahedral crystal field splits the $d$-orbitals into 
$e_g$ and $t_{2g}$ orbitals and the lower energy $e_g$ orbitals 
with two-hold degeneracy form non-bonding states. 
We assume these impurity $d$-levels deep enough in the host VB band. 
Then three $p$-holes and three split-off states are formed.  
The substitution of Ga$^{3+}$ ion by Mn$^{3+}$ can be considered to 
occupy two electrons in $t_{2g}$ orbitals, and this corresponds to 
the doubly occupied split-off states in our model. 
We can interpret this state as $d^5 + 1hole$ configuration. 

\begin{figure}[b]
\centerline{\includegraphics[width=2.2in]{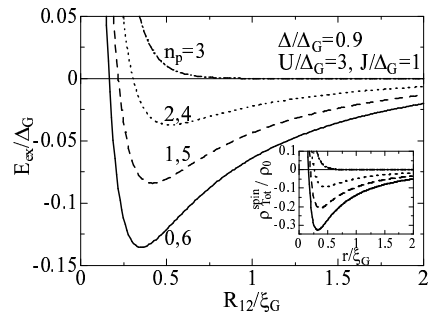}}
\caption{The exchange energy as a function of distance between two impurities 
for various occupation number of the split-off states ($n_p$).
Inset: The spin density distribution of $p$-hole as a function of 
the distance from the impurity site. }
\label{fig2}
\end{figure}

The single impurity case with three degenerate orbitals is analyzed 
in HF approximation with the Coulomb and exchange $J$ terms on an equal footing.    
The spin density of the $p$-holes as a function of 
distance from the impurity site is plotted 
in the inset of Fig.~\ref{fig2}. 
The spin density distribution in negative sign in Fig.~\ref{fig2} represents 
the antiferromagnetic couling with impurity spin. 
Here, $\xi_G=1/\sqrt{2m^*\Delta_G}$ which is about $0.6$nm in GaAs. 
Figure~\ref{fig2} also indicates the spin density of $p$-hole for 
various occupation number of the split-off states ($n_p$). 
The spin density of $p$-hole for $n_p=0,6$ is about three times of that for $n_p=2,4$. 
This is because the split-off states for various $n_p$ 
are nearly degenerate for large $\Delta$ due to the multiple charge state 
by the Haldane-Anderson scenario, then these split-off states give 
the almost same spreading. 

Based on the spin density distribution of $p$-holes around the impurity, 
the exchange interaction mediated by the holes is evaluated 
using the Caroli's formula~\cite{Caroli}, 
which was originally used to analyze the exchange interaction 
of two impurities in metallic host.
We extend this formula to semiconductor host. 
The exchange energy ($E_{ex}$) defined as the energy difference between ferromagnetic 
and antiferromagnetic configurations of two impurity spins are shown in Fig.~\ref{fig2} 
as a function of distance between two impurities. 
Since the  spreading of ferromagnetic coupling over several lattice constants 
shows well agreement with that of $p$-holes, 
these holes plays an essential role for the ferromagnetic exchange interaction. 

\begin{figure}[b]
\centerline{\includegraphics[width=2.2in]{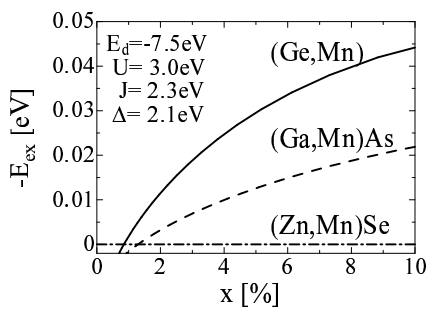}}
\caption{ The exchange energy as a function of impurity concentration 
for various host materials.}
\label{fig3}
\end{figure}

Figure~\ref{fig2} also shows the dependence of the exchange energy on $n_p$. 
Since $n_p$ can be interpreted as the $t_{2g}$ occupation as discussed above, 
the material dependence of the exchange energy 
is examined for various $t_{2g}$ occupation. 
Under the strong $p$-$d$ hybridization and 
deep impurity states in the host VB band, 
the states with occupations $n_p=1,2,$ and $3$ can be considered as hosts Ge, GaAs, 
and ZnSe, respectively. 
The comparison between these Mn doped semiconductors are shown in Fig.~\ref{fig3} 
as a function of impurity concentration. 
Since $T_c$ is proportional to the absolute value of exchange energy 
within HF approximation, 
(Ge,Mn) is expected to have a higher $T_c$ due to a large amplitude of 
$p$-hole spin density around the impurity site. 
On the other hand, (Zn,Mn)Se cannot be  expected higher $T_c$ 
due to no induced $p$-hole only by intrinsic doping. 
This material dependence of the exchange energy 
is roughly similar to the results for the specific doping rate 
by first-principles electronic structure calculation 
done for the periodic supercells~\cite{Butler} and nano-crystals~\cite{Huang}. 
However, these first-principles results~\cite{Butler,Huang} 
have shown the antiferromagnetic ground state for (Zn,Mn)Se.   

\section{Summary}
To summarize this work, the magnetic impurity states and 
magnetic interaction between $d$-impurities are studied 
using extended Haldane-Anderson model in DMS. 
Due to the strong hybridization between $d$-states and $p$-states in the host, 
$p$-holes couple antiferromagnetically to impurity spins and 
the split-off states in the gap are formed. 
When the occupation of the split-off states is different from the number of $p$-holes, 
the spin density of $p$-holes spreads over several lattice constants 
around the impurity site. 
When two impurities are so close to overlap their spin density, 
ferromagnetic interaction is favorable. 
The weakly bound holes around the impurity site 
plays an essential role 
for the ferromagnetic exchange interaction. 
\section*{Acknowledgments}
The part of this work is supported by CREST, IFCAM, and MEXT. 
One of the authors (M.I) thanks the NAREGI project of MEXT Japan. 
\balance
%


\begin{thebibliography}{9}
\bibitem{MaekawaShinjo}
{\em Spin Dependent Transport in Magnetic Nanostructures }, 
edited by S. Maekawa and T. Shinjo 
(Taylor and Francis, London and New York, 2002).

\bibitem{Ohno1}
H. Ohno, {\it et al.},  \Journal{Appl. Phys. Lett.}{69}{363}{1996}.

\bibitem{Matsukura}
F. Matsukura, Y. Ohno and H. Ohno, \Journal{Phys. Rev.} {B57}{R2037}{1998}.

\bibitem{Ohno2}
H. Ohno, {\it et al.}, \Journal{Nature}{408}{944}{2000}.

\bibitem{Ohno3}
D. Chiba, M. Yamanouchi, F. Matsukura, and H. Ohno, \Journal{Science}{301}{943}{2003}.

\bibitem{Okabayashi1}
J. Okabayashi, {\it et al.}, \Journal{Phys. Rev.}{B58}{R4211}{1998}.

\bibitem{Okabayashi2}
J. Okabayashi, {\it et al.}, \Journal{Phys. Rev.}{B59}{R2486}{1999}.

\bibitem{Okabayashi3}
J. Okabayashi, {\it et al.}, \Journal{Phys. Rev.}{B64}{125304}{2001}.

\bibitem{Hirakawa}
K. Hirakawa, {\it et al.}, \Journal{Phys. Rev.}{B65}{193312}{2002}.

\bibitem{Ishikawa}
Y. Ishikawa, {\it et al.}, \Journal{Phys. Rev.}{B65}{233201}{2002}.
 
\bibitem{H&A}
F. D. M. Haldane and P. W. Anderson, \Journal{Phys. Rev.}{B13}{2553}{1976}. 

\bibitem{Caroli}
B. Caroli, \Journal{J. Phys. Chem. Solids}{28}{1427}{1967}. 
\bibitem{Butler}
T. C. Schulthess and W. H. Butler,   \Journal{J. Appl. Phys.}{89}{7021}{2001}.

\bibitem{Huang}
X. Huang, {\it et al.}, \Journal{Phys. Rev. Lett}{94}{236801}{2005}. 

\end{thebibliography}
\end{document}